\documentclass[%
 prd,
 twocolumn,
 superscriptaddress,
 numerical,
 showpacs,
 amsmath,amssymb,
 aps,
nofootinbib,
longbibliography,
 floatfix
]{revtex4-1}

\usepackage{bm}
\usepackage{bbm}
\usepackage{soul}
\usepackage{ulem}
\usepackage{cancel}
\usepackage{braket}
\usepackage{slashed}
\usepackage{verbatim}
\usepackage{graphicx}
\usepackage{hyperref}
\usepackage{combelow} 
\usepackage{mathtools}
\usepackage[usenames,dvipsnames,svgnames,table]{xcolor}

\definecolor{DeepPink}{rgb}{0.796,0.004,0.384}
\definecolor{RoyalBlue}{rgb}{0.02,0.016,0.667}
\definecolor{Teja}{rgb}{0.65,0.20,0.15}
\definecolor{Berry}{rgb}{0.60,0.06,0.34}

\hypersetup{
    colorlinks=true,
    linkcolor=DeepPink,
    urlcolor=cyan,
    citecolor=RoyalBlue,}

\let\oldciteauthor=\citeauthor
\def\citeauthor#1{\hypersetup{citecolor=black}\oldciteauthor{#1}}

\newcommand{\beqn}{\begin{eqnarray}}
\newcommand{\eeqn}{\end{eqnarray}}
\newcommand{\beqs}{\begin{subequations}}
\newcommand{\eeqs}{\end{subequations}\\[-2mm]\noindent}

\definecolor{purple}{rgb}{0.8,0,0.6}
\definecolor{PURPLE}{rgb}{0.8,0,0.6}
\definecolor{orange}{rgb}{1,0.55,0}
\definecolor{limegreen}{rgb}{0.2,0.8,0.2}

\definecolor{battleshipgrey}{rgb}{0.2, 0.52, 0.51}

\allowdisplaybreaks
\begin{document}

\title{
Chiral symmetry breaking and inhomogeneous phases in thermal anti-de Sitter spacetime
}

\author{Sergio Morales Tejera}
\affiliation{Department of Physics, West University of Timi\cb{s}oara, Bd.~Vasile P\^arvan 4, Timi\cb{s}oara 300223, Romania}

\begin{abstract}
We study the spontaneous breaking of chiral symmetry in an AdS spacetime at finite temperature using the quark-meson model. The condensate $\sigma$ is typically inhomogeneous in AdS and is determined from the \textit{differential gap equation}. We demonstrate that there are no free integration constants in the regular solutions to the differential equation and find that the solution to the boundary value problem is unique. We find that chiral symmetry is always broken close to the AdS boundary. We construct the phase diagram of the system as a function of the AdS curvature and temperature. These two parameters have opposing effects: temperature tends to restore chiral symmetry, whereas negative curvature favors its spontaneous breaking. We also consider how the phase diagram is modified when the Hawking-Page phase transition is taken into account.

\end{abstract}

\maketitle

\section{Introduction}\label{sec:intro}

The properties of quantum field theories in curved spacetime backgrounds have received considerable attention, mainly because an accurate description of the physics of our curved universe requires it\footnote{Interestingly, it has also been argued that such studies may be relevant in condensed matter physics, where curvature-related defects or extrinsic curvature can modify the response of materials \cite{Gies:2013dca}.}. In this work, we shall focus on the modification of spontaneous chiral symmetry breaking ($\chi$SB)  on an anti-de Sitter (AdS) background.

The study of the spontaneous breaking of chiral symmetry in curved spaces, usually of constant curvature, began around thirty years ago. Initially, the curvature effects were considered within the Gross-Neveu model \cite{Kanemura:1994rs,Kanemura:1995sx,Inagaki:1995bk} under the assumption of a weakly curved spacetime. Generically, a curvature-induced phase transition was predicted at zero temperature for positively curved spacetimes; negatively curved spaces would remain in the chirally broken phase. A curvature-induced phase transition was also observed in the Thirring model \cite{Geyer:1996yf}. Results on $\chi$SB in curved backgrounds were later rephrased in terms of the NJL model, in which the combined effects of gravity and electromagnetic fields were reported \cite{Gitman:1996mk,Geyer:1996kg,Elizalde:1997aw,Inagaki:1997kz,Elizalde:1998qe}. More recently, the O($N$) linear sigma model in Rindler and AdS spaces has been considered in \cite{Basu:2023bcu}.  The general conclusion stayed for negatively curved spacetimes that no curvature induced phase transition can occur at zero temperature. A physical explanation of this general phenomenon was put forward in Ref. \cite{Gorbar:1999wa}, where it was linked to the effective dynamics of fermion fields in the infrared region becoming effectively 1+1 dimensional. The previous results were soon extended to systems at finite temperature in both positive and negatively curved spaces \cite{Vitale:1998wm,Goyal:2000yx,Huang:2006fk}, as well as to spaces with a different topology \cite{Ishikawa:1996jb,Kim:1997ak,Ebert:2015vua}. Recently, the effect of torsion on $\chi$SB was also addressed \cite{deBrito:2023zxl}.

In recent years, interest has shifted towards more generic curved backgrounds. Notably, the chiral symmetry phase transition has been studied in a variety of black hole backgrounds in asymptotically Minkowski spacetimes \cite{Flachi:2011sx,Quinta:2019hrf,DeMott:2022qux,Tanaka:2026geo}. In these studies, inhomogeneous phases have been constructed in which chiral symmetry is restored close to the black hole event horizon. The effect of gravitational instantons has been considered in \cite{Hamada:2020mug}, and further generic implications of gravitational effects on chiral symmetry can be found in \cite{Flachi:2014jra,Gies:2018jnv}. Curvature-induced phase transitions have also been addressed in the context of holography for the Einstein-scalar theory in \cite{Ghosh:2017big,Kastikainen:2025eys} and for the improved holographic QCD model in \cite{Kiritsis:2025yke}.

Crucially, the order parameter in curved spacetime usually varies with the spacetime coordinates\footnote{Note the exception of the highly symmetric Einstein universe considered in Ref. \cite{Huang:2006fk}, where the condensate remains homogeneous.}. In the weakly curved regime, considered in the early studies of this problem \cite{Kanemura:1994rs,Kanemura:1995sx,Inagaki:1995bk,Geyer:1996yf,Gitman:1996mk,Geyer:1996kg,Elizalde:1997aw,Inagaki:1997kz,Elizalde:1998qe,Goyal:2000yx}, only the linear correction proportional to the Ricci scalar $R$ is considered. Consequently, the condensate $\sigma$ is effectively homogeneous, and the classification of phases proceeds as in flat space. Later on, the weak curvature assumption was dropped, although a homogeneous solution was still imposed on the system, as in Ref.   \cite{Vitale:1998wm}. This approximation is similar to that employed in \cite{Singha:2024tpo,Singha:2025zvh} to obtain homogeneous solutions for rotating matter in flat space. A further step has been taken in the black hole backgrounds \cite{Flachi:2011sx,Quinta:2019hrf,DeMott:2022qux,Tanaka:2026geo} where the inhomogeneous condensate $\sigma$ has been found by solving the Klein-Gordon equation. The same approach was taken in \cite{Morales-Tejera:2025qvh} to characterize the inhomogeneous symmetry breaking pattern for rotating systems in flat space. 

The main objective of this study is to characterize the spontaneous breaking of chiral symmetry in an AdS background, considering its inhomogeneous nature, which is induced by the curved background. To achieve this, we will use the quark meson model and solve the Klein-Gordon equation for the condensate $\sigma$ in this setup. Previous works \cite{Flachi:2011sx,Quinta:2019hrf,DeMott:2022qux} have been obtained have obtained the effective thermodynamic potential through a truncated heat kernel expansion for curved backgrounds and inhomogeneous mass.
In this work, we shall exploit the fact that the fermion condensate in a thermal AdS background has been obtained in closed form in \cite{Ambrus:2017cow}. This condensate is obtained for a constant mass and can be considered a proxy for the behavior of the condensate evaluated with an inhomogeneous mass. We will later discuss the limitations of this approximation.

This paper is organized as follows: In Sec. \ref{sec:model} we present the quark-meson model in an AdS background. In Sec. \ref{sec:finite_effects}, we obtain the inhomogeneous condensate $\sigma$ with two different approaches: (a) assuming that the condensate $\sigma$ is slowly varying, thereby neglecting its gradients (we refer to this as the local density approximations) and (b) solving the Klein-Gordon equation for $\sigma$ in the AdS background. We characterise the phase of the system based on the value of $\sigma$ at the center of AdS. In Sec. \ref{sec:HP} consider the modification of the phase diagram when the Hawking-Page transition is taken into account. In this section, we evaluate the fermion condensate using the local thermal equilibrium approximation. In Sec. \ref{sec:holo} we comment on the generic properties of the holographic dual to the quark meson model in AdS. We conclude in Sec. \ref{sec:conc} by summarizing our findings and outlining future avenues of research. 

\section{Quark-meson model in AdS}\label{sec:model}

In order to study the curvature effects on $\chi$SB, we employ the quark-meson model with two flavours and assume minimal coupling to gravity. The classical action of the model, with a generic background metric $g^{\mu\nu}$, is 
\begin{multline}
    S =\int_{\mathcal{M}} d^4x\sqrt{-g}\left[\dfrac{1}{2\kappa^2}\left(R-2\Lambda\right) + \mathcal{L}_{\rm QM}\right] \\+ \dfrac{1}{\kappa^2}\int_{\partial\mathcal{M}}d^3x\sqrt{-\gamma}K\,,
\end{multline}
where $\kappa^2$ and $\Lambda$ are the Newton and cosmological constants respectively. The Ricci scalar is denoted $R$, while $K$ represents the trace of the extrinsic curvature at the boundary of the manifold $\partial \mathcal{M}$ and $\gamma^{\mu\nu}$ is the induced metric at the boundary. We shall work in the probe limit, where the backreaction of matter onto the geometry is neglected. We comment on the shortcomings of this limit below. The Lagrangian for the quark meson model is given by
\begin{multline}
     \mathcal{L}_{\rm LSM}  = \dfrac{1}{2}\partial_\mu \sigma\partial^\mu\sigma + \dfrac{1}{2}\partial_\mu \vec{\pi}\partial^\mu\vec{\pi} + U(\sigma,\vec{\pi}) + \\\overline{\psi}\left(\dfrac{i}{2}\overleftrightarrow{\slashed{\nabla}} -g (\sigma + i \gamma^5 \vec{\tau} \cdot \vec{\pi})\right)\psi
\end{multline}
where the mesonic potential is 
\begin{equation}
    U(\sigma,\vec{\pi}) = \frac{\lambda}{4}(\sigma^2+\vec{\pi}^2-v^2)^2 - h\sigma\,.
\end{equation}
 The parameters of the model are $(\lambda,v,h,g)$, which are customarily matched to the vacuum properties of the model in flat spacetime. In particular, the parameters are set to reproduce the values of the pion decay constant $f_\pi \simeq 93 \ \rm MeV$, and the masses of the pion $m_\pi \simeq 138 \ \rm MeV$, the sigma meson $m_\sigma \simeq 600 \ \rm MeV$, and the constituent quarks $m_q \simeq 307\ \rm MeV$. In this case, one obtains
 \begin{equation}\label{eq:params}
     \lambda = 19.71\,,\ g=3.3\,,\ v=87.75\,\ {\rm MeV}\,,\ h=1.77\cdot 10^6 \,\rm MeV^3\,.
 \end{equation}
We shall address the thermodynamic properties of the model, which are encoded in the grand canonical potential
\begin{equation}
    \Phi = -\dfrac{1}{\beta_0}\ln\mathcal{Z}\,,
\end{equation}
where $\beta_0$ is the inverse temperature and $\mathcal{Z}$ is the partition function, which we evaluate in the mean-field approximation for the pions and the condensate $\sigma$, in the background metric $g^{\mu\nu}$:
\begin{equation}
    \mathcal{Z}[\sigma,\vec{\pi},g^{\mu\nu}] = \int \mathcal{D\psi D\overline\psi}\ e^{iS[\sigma,\vec{\pi},\psi,\overline{\psi},g^{\mu\nu}]}\,.
\end{equation}
The gap equations that follows form requiring the extremization of the grand canonical potential with respect to the mesonic fields are 
\begin{equation}\label{eq_gap}
    \square \sigma - \lambda(\sigma^2+\vec{\pi}^2-v^2)\sigma   =  g\langle \overline\psi\psi\rangle-h\,.
\end{equation}
\begin{equation}\label{eq_pions}
    \square \vec{\pi}- \lambda(\sigma^2+\vec{\pi}^2-v^2)\vec{\pi}  =  -g\langle -i\overline\psi\gamma^5\vec{\tau}\psi\rangle\,,
\end{equation}
where the expectation value of the scalar and pseudoscalar condensates is taken at finite temperature on the AdS background.

We choose to work in $3+1$-dimensional AdS space written in global coordinates with a compactified radial direction. Then, the line element is written as
\begin{equation}\label{eq:AdS_metric}
    ds^2 = \dfrac{\ell^2}{\cos^2 r}\left(-dt^2 + dr^2 + \sin^2r  d\Omega_2^2\right)\,,\quad r\in[0,\pi/2)
\end{equation}
where all coordinates are dimensionless. The boundary of AdS is located at $r=\pi/2$, while the center of AdS is at $r=0$. The AdS scale $\ell$ controls the curvature of the spacetime. In particular, the Ricci scalar is given by
\begin{equation}
    R = 4\Lambda=-\frac{12}{\ell^2}\,.
\end{equation}

The thermal expectation value of the fermion scalar and pseudoscalar condensates in an AdS background at finite temperature has been evaluated in Refs. \cite{Ambrus:2017cow,Ambrus:2021eod} for a fermion species of constant mass $M$, and we quote here the result:
\begin{multline}\label{eq:scalar_cond}
    \langle\overline\psi\psi \rangle_{T,\ell} = \dfrac{\Gamma_k}{2\pi^2\ell^2}\sum_{j=1}^\infty(-1)^{j+1}\zeta_j^{2+k}\cosh{\dfrac{j\beta_0}{2\ell}} \\\times {_2}F_1 (1+k,2+k;1+2k;-\zeta_j) 
\end{multline}
\begin{equation}\label{eq:pseudoscalar_cond}
    \langle\overline{\psi}i\gamma^5\psi\rangle_{T,\ell} =0
\end{equation}
where 
\begin{equation}
    \zeta_j = \dfrac{\cos^2r}{\sinh^2(j\beta_0/2\ell)};\quad \Gamma_k = \dfrac{\Gamma(2+k)\Gamma\left(\frac{1}{2}\right)}{4^k\Gamma\left(\frac{1}{2}+k\right)},
\end{equation}
with $k = M\ell$.
In our case, the mass of the fermion is effectively given by the expectation value of the condensate $\sigma$, $M=g\sigma$. Note that we work with two light flavors, up and down, and with isospin symmetry, which implies $m_u=m_d$. The right hand side of Eq. \eqref{eq_pions} is identically zero and the mean-value of the pions is trivial $\vec{\pi} =0$. The vacuum contribution to the fermion condensate is given by
\begin{multline}\label{eq:cond_vacuum}
    \langle\overline\psi\psi\rangle_{\rm vac} = -\dfrac{1}{4\pi^2\ell^3}\left(1+\dfrac{k}{6}-k^2-k^3\right.\\\left. -2k(1-k^2)\left[\psi(1+k)+\dfrac{1}{2} - \ln k\right]\right)\,.
\end{multline}
Therefore, the fermion condensate that serves as the source of the gap equation \eqref{eq_gap} is given by
\begin{equation}\label{eq_fermion_total}
    \langle\overline\psi\psi\rangle = N_f N_c \left[\langle\overline\psi\psi\rangle_{\rm vac}+\langle\overline\psi\psi\rangle_{T,\ell}\right]\,,
\end{equation}
where we have introduced the flavor and color degeneracy factors. Note that the thermal part of the fermion condensate \eqref{eq:scalar_cond} depends on the AdS radial coordinate $r$, and so will the generated effective mass.

\section{Finite temperature and curvature effects in AdS}\label{sec:finite_effects}

In this section, we will solve the gap equation \eqref{eq_gap} and discuss the combined effect of temperature and negative curvature on the mass gap and phase of the system. The curvature of spacetime enters the gap equation explicitly through the fermion condensate \eqref{eq_fermion_total} and through the gradient terms $\square \sigma $. In order to disentangle the effect of each of them, we shall first solve the gap equation in the absence of gradient terms in Sec. \ref{sec:LDA}. The condensate $\sigma$ is then a local function of the radial coordinate $r$ and we refer to this approach as the local density approximation (LDA). Then, in Sec. \ref{sec:LI}, we include the gradient terms and solve the gap equation in its differential form. 

\subsection{Local density approximation (LDA)}\label{sec:LDA}

\begin{figure}
    \centering
    \includegraphics[width=0.95\linewidth]{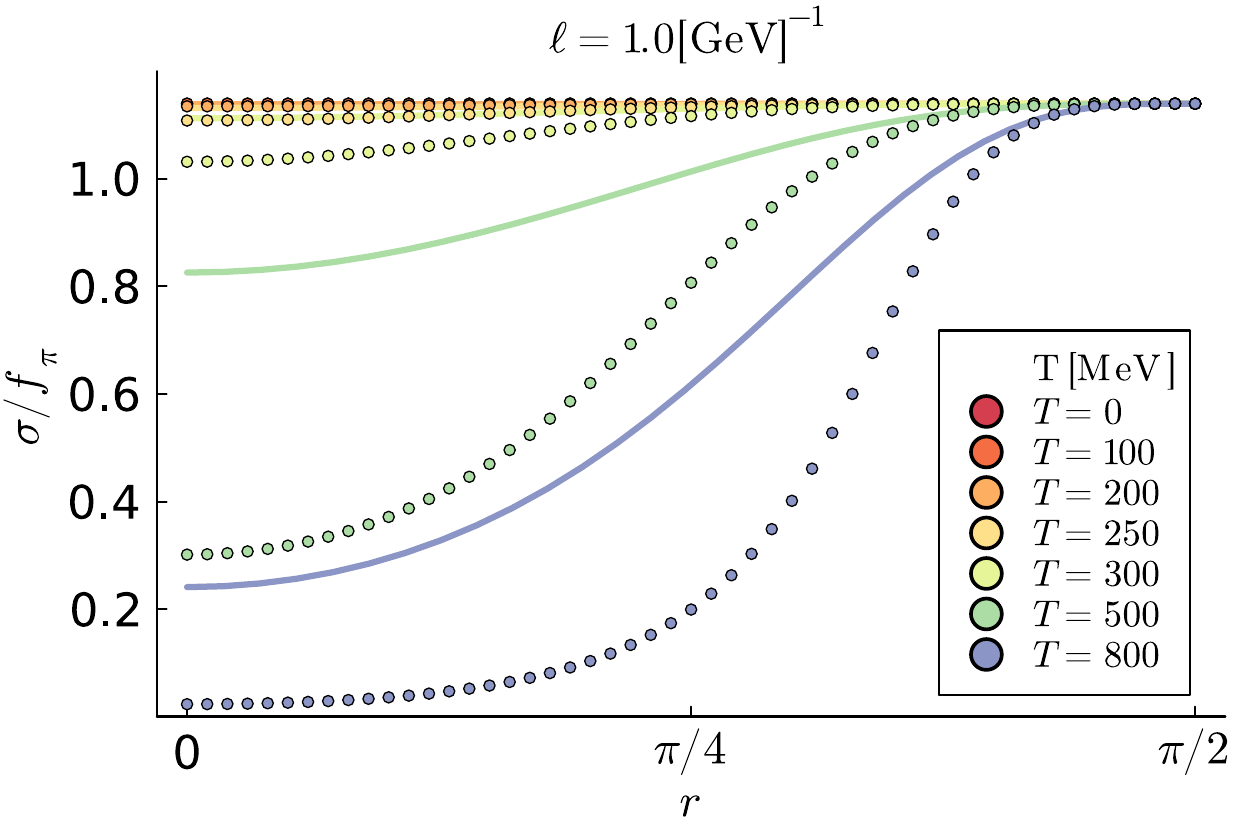}
     \includegraphics[width=0.95\linewidth]{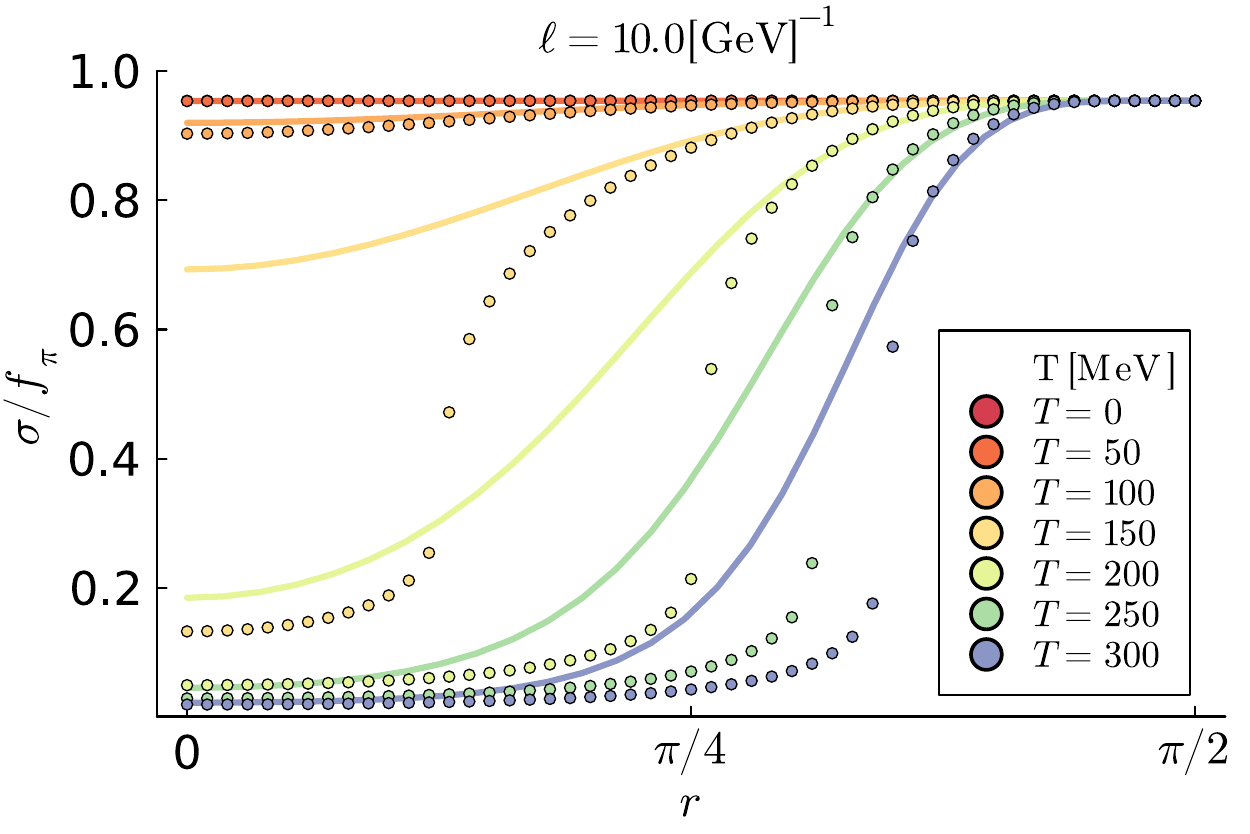}
    \caption{Dimensionless condensate $\sigma/f_\pi$ as a function of the dimensionless radial distance to the center of AdS: $r$. The value of $\sigma$ is obtained as a solution of the gap equation \eqref{eq_gap} under the local density approximation (scattered points) and solving the boundary value problem (solid lines). The results are shown at two different AdS scales $\ell$ and several temperatures $T$. }
    \label{fig:s_rbar_l1}
\end{figure}

In general, the fermion condensate \eqref{eq_fermion_total} depends on the radial coordinate $r$, and so does the condensate $\sigma$ solving the gap equation \eqref{eq_gap}. The gradients are controlled by $\ell$ [see Eq. \eqref{eq:gap_difeq_AdS}], so when the AdS scale is very large, $\ell\gg 1$, the contribution of the gradients to the gap equation is parametrically small, and we expect that the gradients can be safely neglected in this limit. Similarly, when the temperature is much smaller than the AdS curvature $T\ell\ll1$, the thermal contribution to the fermion condensate \eqref{eq:scalar_cond} (responsible for the inhomogeneity of $\sigma$) vanishes exponentially, while the vacuum contribution remains finite. Therefore, in this limit, the condensate $\sigma$ is approximately homogeneous and the gradients can also be neglected: $\square \sigma \simeq 0$. In this section, we solve the gap equation under the approximation that $\square \sigma \simeq 0$:
\begin{equation}\label{eq_algebraicgap}
 \lambda(\sigma^2-v^2)\sigma   = h - g\langle \overline\psi\psi\rangle\,.
\end{equation}
The value of the condensate $\sigma$ as a function of the dimensionless radial distance $r$ is shown in Fig. \ref{fig:s_rbar_l1} at several temperatures $T$ and different AdS scales $\ell$. The \textit{scattered} points correspond to the solution to the gap equation under the LDA. The value of the condensate $\sigma$ at the boundary is independent of the temperature. Indeed, at $r=\pi/2$, the thermal contribution to the fermion condensate vanishes, and the solution to the algebraic gap equation \eqref{eq_algebraicgap} is controlled by the (temperature-independent) vacuum contribution to the fermion condensate. As we depart from the boundary, the condensate decreases until it reaches a certain value at the center $r=0$, which does depend on the temperature of the system. We observe that the condensate $\sigma$ is monotonic as a function of the radial coordinate $r$.  The system is either in a chirally broken phase as a whole, or in a mixed phase, where chiral symmetry is broken close to the boundary and restored close to the center of AdS. Generically, we observe that increasing the temperature results into the restoration of chiral symmetry close to the center of AdS, while increasing the curvature prevents the restoration of the symmetry. This observation is in agreement with the observed gravitational catalysis on negatively curved spaces \cite{Kanemura:1994rs,Kanemura:1995sx,Inagaki:1995bk}. 

\begin{figure}
    \centering
    \includegraphics[width=0.95\linewidth]{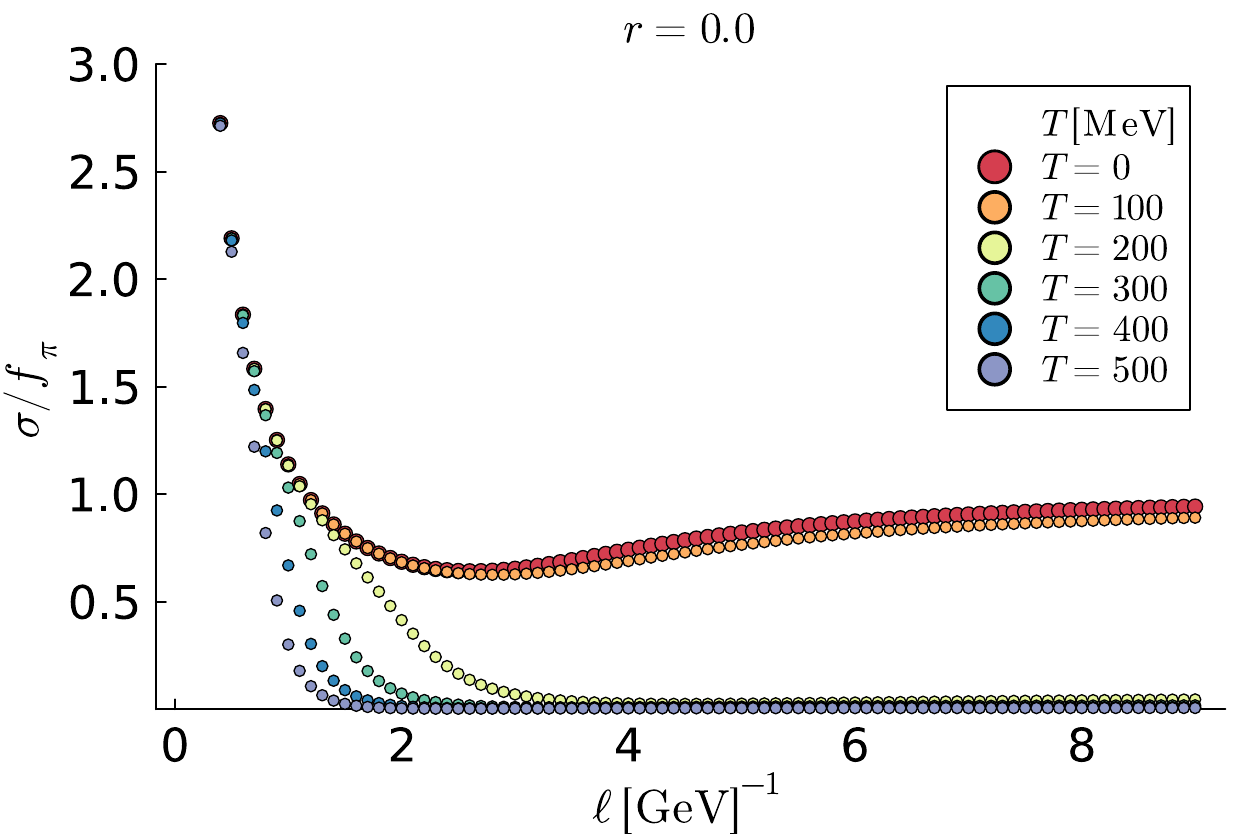}
    \caption{Dimensionless condensate $\sigma/f_\pi$ at the center of AdS, $r=0$, as a function of the AdS scale $\ell$ for different temperatures in the LDA, see Eq. \eqref{eq_algebraicgap}. The curve for $T=0$ coincides with the value of $\sigma/f_\pi$ at the boundary of AdS for \textit{any} temperature.}
    \label{fig:s_l_rbar0}
\end{figure}

In Fig. \ref{fig:s_l_rbar0}, we show how the condensate $\sigma$ at the center of AdS is affected by the AdS scale $\ell$. Interestingly, the condensate $\sigma$ can feature a non-monotonic behavior, and it is bounded from below but not from above. In particular, the effective fermion mass $g\sigma$ diverges as the curvature diverges ($\ell \to 0$). In this regime, we expect that the backreaction of matter onto the geometry is sizeable and cannot be neglected. 

The value of the condensate $\sigma$ at the center of AdS for $T=0$ (red points in Fig. \ref{fig:s_l_rbar0}) is the same for the whole spacetime, since the fermion condensate is homogeneous at vanishing temperature. 

The red points in Fig. \ref{fig:s_l_rbar0} also represent the boundary value of $\sigma$ for \textit{any} temperature. This is a consequence of the fact that the fermion condensate vanishes at the boundary together with the fact that at $T=0$ the condensate is homogeneous. It can then be seen in Fig. \ref{fig:s_l_rbar0} that the minimum value of the condensate $\sigma$ at the boundary of AdS is $\sigma_{\rm min} \simeq 0.64f_\pi $ at $\ell\simeq 2.6$ GeV$^{-1}$. We can therefore conclude that, close to the boundary of AdS, the system is always in the chirally broken phase for the LDA. Whether the system is globally in the chirally restored phase, or in a mixed phase, is determined by the value of the condensate at the center of AdS. 

\subsection{Differential gap equation}\label{sec:LI}

In this section, we solve the equation of motion for the condensate $\sigma$ \eqref{eq_gap} taking into account the local gradients $\square \sigma\neq0$. In the AdS background \eqref{eq:AdS_metric}, the equation of motion becomes 
\begin{equation}
    \dfrac{1}{\ell^2}\left(\cos^2r\partial_r^2+2\cot r\partial_r\right)\sigma -\lambda(\sigma^2-v^2)\sigma +h - g\langle\overline{\psi}\psi\rangle = 0\,.
\end{equation}
It is convenient to change variables to $u =\cos r$, in which case the boundary is mapped to $u=0$ while the center of AdS is at $u=1$. Then, the previous equation can be rewritten as
\begin{align}\label{eq:gap_difeq_AdS}
   & \dfrac{1}{\ell^2}\left[u^2(1-u^2)\partial_u^2-u(2+u^2)\partial_u\right]\sigma =\nonumber \\ &=\lambda(\sigma^2-v^2)\sigma -h + g\langle\overline{\psi}\psi\rangle\,.
\end{align}
The boundary and the center of AdS are singular points of the differential equation above. In order to understand the global structure of the solutions of Eq. \eqref{eq:gap_difeq_AdS}, we study the asymptotic behaviour of the condensate $\sigma$ around the two singular points.

\subsubsection{Asymptotic solution around the center of AdS}

At the center of AdS, $u=1$, the fermion condensate is generically non-vanishing and we can expect that $\sigma$ attains a finite value $\sigma_0$. As a first step to find the asymptotic solution, we Taylor expand $\sigma$:
\begin{equation}\label{eq:sigma_Taylor}
    \sigma_{\rm reg.}(u) = \sum_{n=0}^\infty \dfrac{\sigma_n}{n!}(1-u)^n\,,
\end{equation}
where the subscript 'reg.' denotes the regular part of $\sigma$. In this case, the fermion condensate given in Eq. \eqref{eq_fermion_total} also admits a taylor expansion, which we write schematically as
\begin{equation}\label{eq:FC_Taylorr}
    \langle\overline{\psi}\psi\rangle_{\rm reg.} = \sum_{n=0}^\infty \dfrac{{\rm FC}_n}{n!}(1-u)^n\,,
\end{equation}
where the coefficients FC$_n$ depend on $\sigma_{m\leq n}$.
We substitute Eqs. \eqref{eq:sigma_Taylor} and \eqref{eq:FC_Taylorr} into the differential gap equation \eqref{eq:gap_difeq_AdS} and solve it perturbatively around $u=1$. The first few coefficients for $\sigma$ are found to be 
\begin{equation}
    \sigma_1 = -\dfrac{\ell^2}{3}\left[h-g {\rm FC_0} +\lambda(v^2-\sigma_0^2)\sigma_0 \right]\,,
\end{equation}
\begin{align}
    \sigma_2 = \dfrac{g\ell^2}{5}{\rm FC_1} + \sigma_1 -\dfrac{\lambda\ell^2}{5}(v^2-3\sigma_0^2)\sigma_1\,.
\end{align}
Note that there is a single undetermined parameter of the solution, $\sigma_0$, which is an integration constant. The differential equation \eqref{eq:gap_difeq_AdS} is of second order, and there should be two integration constants. In order to extract the second integration constant, we study small fluctuations around the Taylor expanded solution of Eq. \eqref{eq:sigma_Taylor}, that is we write $\sigma = \sigma_{\rm reg.} + \delta\sigma$. We assume that $\delta\sigma$ is small, which is equivalent to assuming that the missing integration constant is small, and linearize the differential equation. Note that this further induces a modification in the fermion condensate as 
\begin{multline}
    \langle\overline\psi\psi\rangle_{\rm reg.}\to \langle\overline\psi\psi\rangle_{\rm reg.} + \langle\overline\psi\psi\overline \psi \psi\rangle(\sigma_{\rm reg.})\delta \sigma\,,\\ \rm with \quad \langle\overline\psi\psi\overline \psi \psi\rangle=\dfrac{\delta  \langle\overline\psi\psi\rangle}{\delta\sigma}\,.
\end{multline}
The four-point correlation function is evaluated at $\sigma_{\rm reg.}$ and also admits a Taylor expansion close to the center of AdS:
\begin{equation}\label{eq:FC_Taylor}
    \langle\overline{\psi}\psi\overline{\psi}\psi\rangle_{\rm reg.} = \sum_{n=0}^\infty \dfrac{{\rm FC}_n'}{n!}(1-u)^n\,.
\end{equation}
The linearized equation for $\delta \sigma$ is given by
\begin{align}\label{eq:linear_sigma}
     \dfrac{1}{\ell^2}\left[u^2(1-u^2)\partial_u^2-u(2+u^2)\partial_u\right]\delta\sigma \nonumber \\ +\left[\lambda v^2 -3\lambda \sigma_{\rm reg.}^2  - g\langle\overline{\psi}\psi\overline{\psi}\psi\rangle_{\rm reg.}\right]\delta\sigma = 0\,.
\end{align}
The function $\delta \sigma$ that solves \eqref{eq:linear_sigma} should contain the non-analytical part of $\sigma$. For this reason, we try a Frobenius like ansatz
\begin{equation}
    \delta\sigma = (1-u)^k\sum_{n=0}^\infty \delta\sigma_n (1-u)^n\,,
\end{equation}
where the power $k$ is in general not an integer value, and it has to be determined by the equations of motion. We substitute the previous ansatz into Eq. \eqref{eq:linear_sigma}. The leading order contribution is as $u\to 1$ is given by
\begin{equation}
    -\dfrac{\delta\sigma_0}{\ell^2}  (1-u)^{k-1}\left[k(1+2k) + O(1-u)\right] = 0,,
\end{equation}
from which we find that $k=0$ or $k=-1/2$. The first possibility, $k=0$, corresponds to the integration constant $\sigma_0$ and is already included in $\sigma_{\rm reg.}$. The second posibility, $k=-1/2$, would imply that $\delta\sigma = \delta\sigma_0/\sqrt{1-u}\gg\sigma_{\rm reg.}$, contradicting the assumption that $\delta \sigma$ is a small perturbation around $\sigma_{\rm reg.}$. Consequently, we must discard the possibility that $k=-1/2$ and we learn that the local solution to the differential equation \eqref{eq:gap_difeq_AdS} around the center of AdS ($u=1$) has a single integration constant: $\sigma_0$.\footnote{Strictly speaking, one could try a Frobenious-like ansatz for $\sigma$ from the beginning, and find $k=-1/2$. In any case, this would give a divergent $\sigma$ at the center of AdS, which is unphysical and would have to be discarded.}

\subsubsection{Asymptotic solution around the boundary of AdS}

At the boundary of AdS, $u=0$, the thermal contribution to the fermion condensate \eqref{eq:scalar_cond} vanishes, while the vacuum part \eqref{eq:cond_vacuum} remains finite for finite $\ell$. Consequently, we expect that the condensate $\sigma$ attains a finite value at the boundary. Similarly to the previous subsection, we find the asymptotic form of $\sigma$ close to the critical point by splitting $\sigma = \sigma_{\rm reg.}+\delta\sigma$, where $\sigma_{\rm reg.}$ contains the regular part of $\sigma$ that can be expanded in a taylor series \eqref{eq:FC_Taylor}, while $\delta \sigma\ll\sigma_{\rm reg.}$ contains the non-analytical contribution. Close to the boundary, located at $u=0$, we expand
\begin{equation}\label{eq:Taylor_bdy_sigma}
    \sigma_{\rm reg.}^{\rm bdy} = \sum_{n=0}^{\infty}\dfrac{\sigma_{n,\rm bdy}}{n!}u^n\,.
\end{equation}
In this case, the vacuum contribution to the fermion condensate \eqref{eq:cond_vacuum} also admits a Taylor expansion for $\sigma_{0,\rm bdy}\neq0$, while the thermal contribution, given in Eq. \eqref{eq:scalar_cond}, vanishes with a (generically) non-integer power: $u^{4+2 g\sigma \ell}$. Assuming that $\sigma_{0,\rm bdy}>0$, this means that we can neglect the thermal contribution of the fermion condensate close to the boundary at least up to $O(u^4)$. We expand the vacuum contribution to $\langle\overline{\psi}\psi \rangle$ as 
\begin{equation}\label{eq:Taylor_bdy_fermion}
     \langle\overline{\psi}\psi\rangle_{\rm vac.}^{\rm reg.} = \sum_{n=0}^\infty \dfrac{{\rm FC}_n^{\rm vac.}}{n!}u^n\,,
\end{equation}
where the coefficients ${\rm FC}_n^{\rm vac.}$ depend on $\sigma^{\rm bdy}_{m\leq n}$. Substituting the Taylor expansions \eqref{eq:Taylor_bdy_sigma} and \eqref{eq:Taylor_bdy_fermion} into the differential equation \eqref{eq:gap_difeq_AdS}, we find that all the coefficients $\sigma_n$ are fixed. The first few coefficients are determined by the following equations:
\begin{equation}\label{eq:bdygap}
    h-g{\rm FC}_0^{\rm vac} + \lambda(v^2-\sigma_{0,\rm bdy}^2)\sigma_{0,\rm bdy} = 0
\end{equation}
\begin{equation}
    \left[\lambda(v^2-3\sigma_{0,\rm bdy}^2)-\dfrac{2}{\ell^2}\right]\sigma_{1,\rm bdy} - g{\rm FC}_1^{\rm vac.} =0\,.
\end{equation}
In particular, the value of the condensate at the boundary $\sigma_{0,\rm bdy}$ is the same as the one obtained in the LDA, since Eq. \eqref{eq:bdygap} is the algebraic gap equation \eqref{eq_algebraicgap} evaluated at the boundary $u=0$. Recall that the thermal contribution to the fermion condensate vanishes at the boundary and the value of $\sigma$ at the boundary is given by the red points in Fig. \ref{fig:s_l_rbar0}. We conclude that chiral symmetry is locally broken close to the boundary regardless of the temperature.

We observe that the asymptotic form of $\sigma$ close to the boundary of AdS does not contain any integration constant in its regular part. We now proceed to asymptotically solve the differential equation for $\delta\sigma$ given in Eq. \eqref{eq:linear_sigma}. We neglect again the contributions from the thermal part of the fermion condensate, which are relevant only at higher orders in the AdS radial coordinate $u$. We try a Frobenious-like ansatz for $\delta \sigma$ while we Taylor expand the four-point function $\langle\overline\psi\psi\overline\psi\psi\rangle_{\rm vac.}^{\rm reg.}$:
\begin{equation}
    \delta\sigma_{\rm bdy} = u^\Delta\sum_{n=0}^\infty \delta\sigma_{n,\rm bdy} u^n\,,\quad \langle\overline\psi\psi\overline\psi\psi\rangle_{\rm vac.}^{\rm reg.} = \sum_{n=0}^\infty {\rm FC'}_n^{\rm vac.}u^n\,.
\end{equation}
The leading contribution to the differential equation \eqref{eq:linear_sigma} is given by 
\begin{equation}\label{eq:leading_linear}
    \delta \sigma_{0,\rm bdy} u^\Delta\left(\dfrac{\Delta(\Delta-3)}{\ell^2}- m_{\sigma,\rm eff.}^2\right) + O(u^{\Delta+1})=0\,,
\end{equation}
where $m_{\sigma,\rm eff.}$ is the effective mass of the scalar field $\sigma$ at the boundary, and it is given by
\begin{equation}
    m_{\sigma,\rm eff.}^2 = \lambda(3\sigma_{0,\rm bdy}^2-v^2) + g{\rm FC'}_0^{\rm vac.}\,.
\end{equation}
Therefore, we find two possible solutions for $\delta \sigma$:
\begin{equation}\label{eq:dsigma}
    \delta\sigma_{\rm bdy} = \mathcal{C}_+ u^{\Delta_+}[1+O(u)] + \mathcal{C_-}u^{\Delta_-}[1+O(u)]\,,
\end{equation}
where $\Delta_{\pm}$ are the two solutions to Eq. \eqref{eq:leading_linear}:
\begin{equation}\label{eq:delta_pm}
    \Delta_{\pm} = \dfrac{1}{2}\left(3\pm\sqrt{9 + 4 m_{\sigma,\rm eff.}^2\ell^2} \right)\,.
\end{equation}
\begin{figure}
    \centering
    \includegraphics[width=0.95\linewidth]{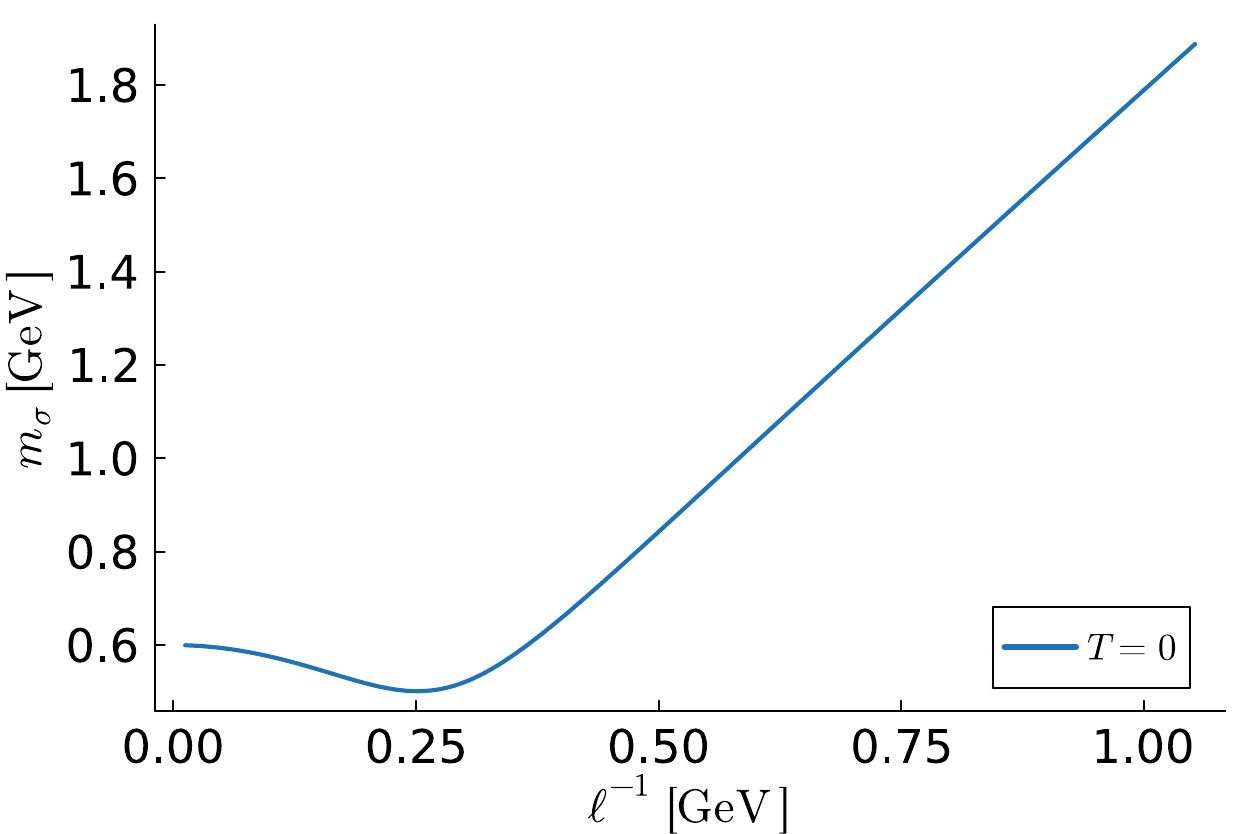}\\
    \includegraphics[width=0.95\linewidth]{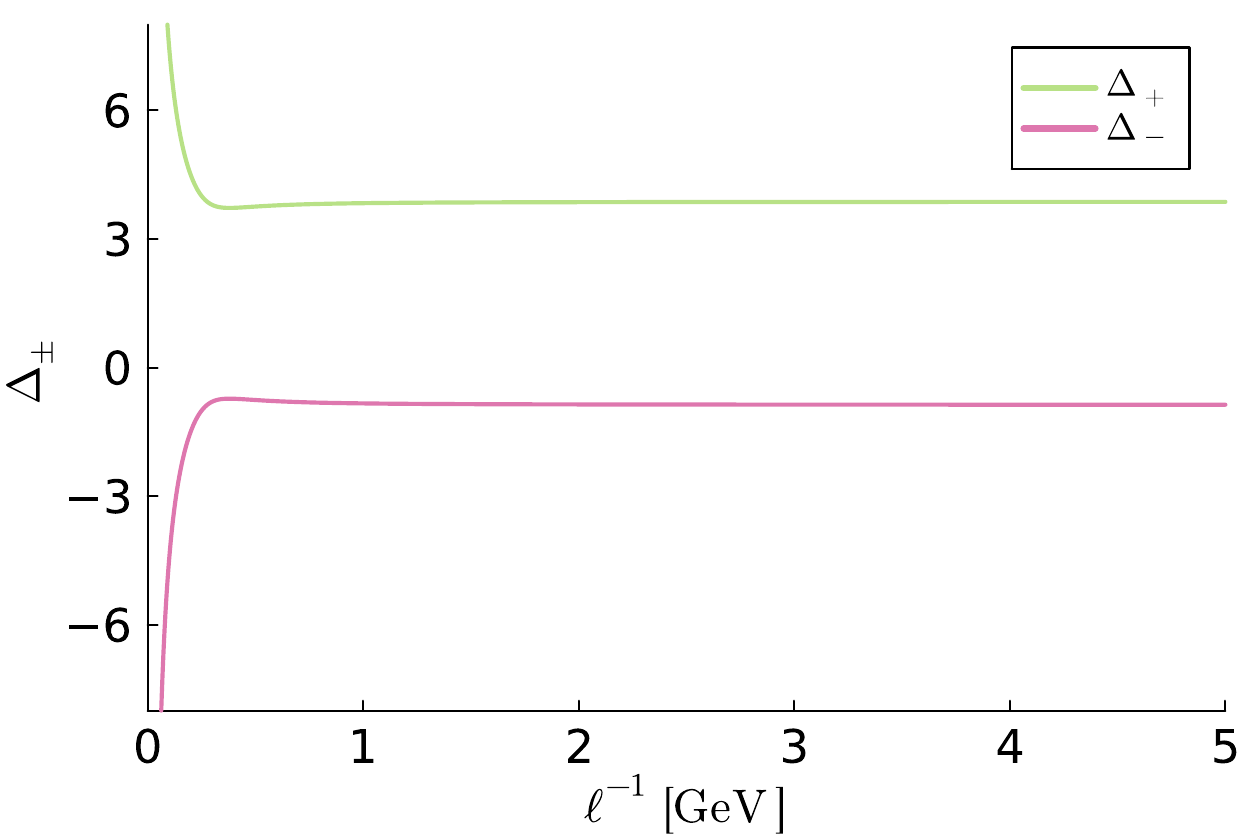}
    \caption{(Top panel) Effective mass of the condensate $\sigma$ at vanishing temperature as a function of the inverse AdS scale $\ell^{-1}$. This coincides with the effective mass close to the boundary. (Bottom panel) Parameters $\Delta_{\pm}$ defined in Eq. \eqref{eq:delta_pm} controlling the non-analytic asymptotic solution of $\sigma$ close to the boundary given in Eq. \eqref{eq:dsigma}. }
    \label{fig:m_eff_Deltas}
\end{figure}
In the top panel of Fig. \ref{fig:m_eff_Deltas} we show the effective mass of $\sigma$ at the boundary, which coincides with its effective mass at zero temperature, as a function of the AdS curvature $R\sim \ell^{-1}$. We observe that $m_{\sigma}^{\rm eff. }$ slightly decreases for small $\ell^{-1}$ while it increases linearly for large curvature. The values of $\Delta_{\pm}$ defined in Eq. \eqref{eq:delta_pm} are displayed in the bottom panel of Fig. \ref{fig:m_eff_Deltas}. Note that $\Delta_+>0$ while $\Delta_-<0$ and, as a consequence, the solution in Eq. \eqref{eq:dsigma} proportional to $u^{\Delta_-}$ is not subleading with respect to $\sigma_{\rm reg.}$. Therefore, we have to set $\mathcal{C}_-=0$ for consistency. We conclude that the asymptotic solution to the differential equation \eqref{eq:gap_difeq_AdS} around the boundary of AdS has a single integration constant: $\mathcal{C}_+$.

\subsubsection{Boundary value problem}\label{sec:bdy_problem}

The results from the previous two subsections allow us to set the boundary value problem for the differential gap equation \eqref{eq:gap_difeq_AdS}. In particular, a solution that extends over the whole AdS spacetime $u\in(0,1]$ has a single integration constant close to the boundary, $\mathcal{C}_+$, which has to be tuned such that the solution is regular at the center. The boundary value problem can be set as Eq. \eqref{eq:gap_difeq_AdS} together with the two conditions that (i) $\sigma$ is finite at the boundary ($u=0$) and (ii) $\sigma$ is finite at the center ($u=1$).

Generically, there will only be a discrete set of integration constants $\{\sigma_{0,i}\}$ that yield a finite condensate at the center. We find that, restricting to positive values of the condensate $\sigma_0>0$, there is a unique $\sigma_0$ that solves the boundary value problem.

In Fig. \ref{fig:s_rbar_l1} we show the solution to the boundary value problem (solid lines) for several temperatures and two values of the AdS scale $\ell$. We confirm our expectations of Sec. \ref{sec:LDA} that the LDA is sufficiently close to the local inhomogeneous solution in the cases where $\ell$ is sufficiently large or when $\ell T\ll 1$. Additionally, we observe that the gradients tend to increase the overall value of $\sigma$, favoring $\chi$SB. We conclude that gradients in AdS also act as a catalyzer for $\chi$SB. This is opposite to the effect of gradients in the rotating system in flat space reported in Ref. \cite{Morales-Tejera:2025qvh}.

Finally,  Eq. \eqref{eq:bdygap} shows that the value of $\sigma$ at the boundary is the same as in the LDA approach, and the system is \textit{always} in the chirally broken phase close to the boundary. Similarly to the LDA, the condensate $\sigma$ is a monotonic function of the AdS radial coordinate, decreasing as it departs form the boundary. Therefore, the phase of the system is either globally chirally broken or inhomogeneous: chirally broken close to the boundary and chirally restored close to the center. The phase of the system is ultimately determined by the value of the condensate at the center: $\sigma_0$.

\subsection{Condensate $\sigma$ and phase diagrams}

In Secs. \ref{sec:LDA} and \ref{sec:LI} we have shown that the phase of the system is determined by the value of the condensate $\sigma $ at the center. In this section, we focus on the behaviour of $\sigma(\bar{r}=0)$ in the two approaches.  

\begin{figure}
    \centering
    \includegraphics[width=0.95\linewidth]{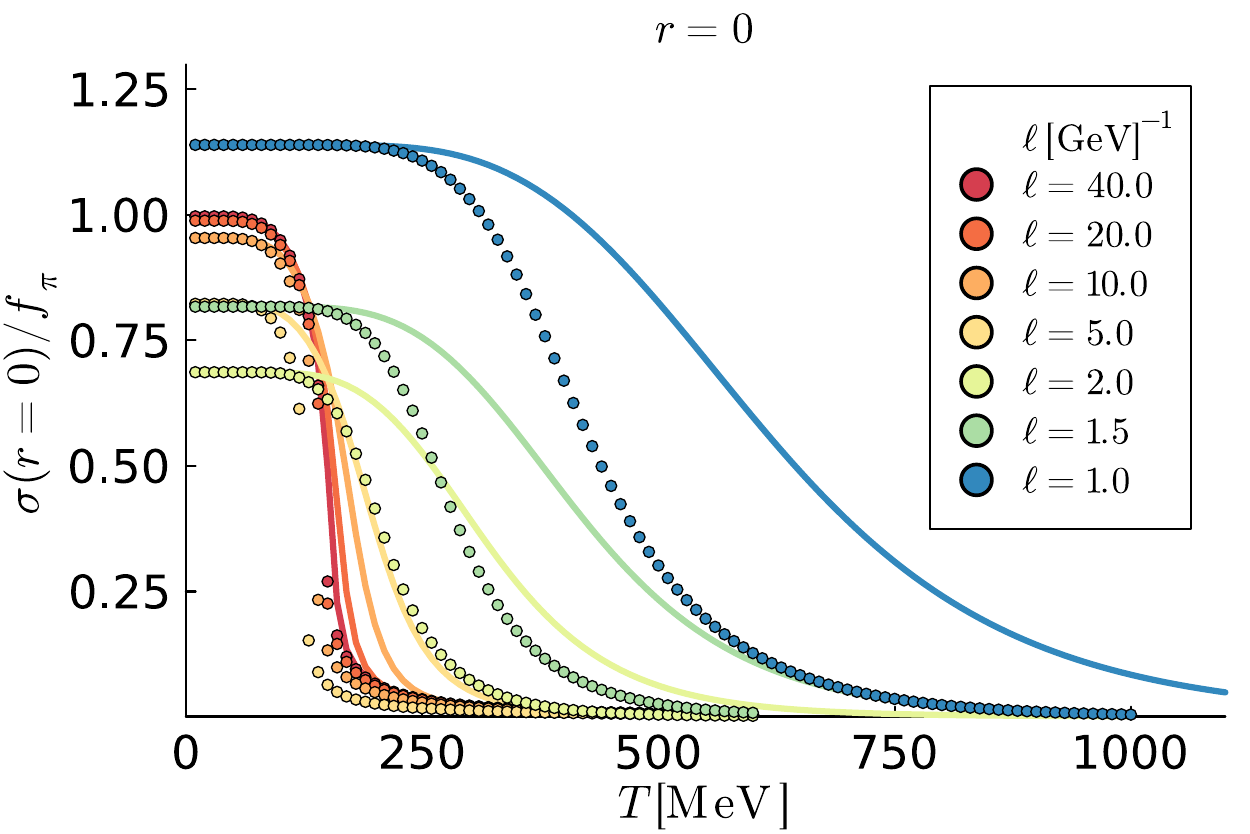}
    \caption{Dimensionless condensate $\sigma/f_\pi$ at the center of AdS ($ r = 0$) as a function of the temperature for different values of $\ell$. The value of $\sigma$ is obtained as a solution of the gap equation \eqref{eq_gap} under the local density approximation (scattered points) and solving the boundary value problem (solid lines).}
    \label{fig:s_T_rbar0}
\end{figure}

In Fig. \ref{fig:s_T_rbar0} we show the condensate $\sigma$ at the center as a function of temperature for several AdS scales $\ell$. Solid points correspond to the solution of the gap equation in the LDA while solid lines are obtained from the solution to the boundary value problem. In both cases, the condensate vanishes for sufficiently large temperatures, while at zero temperature it is bounded below by $\sigma_{\rm min}\simeq 0.64 f_\pi$ (see Fig. \ref{fig:s_l_rbar0}). We define the phase transition temperature as the point where the rate of change in $\sigma$ as a function of the temperature $T$ is maximum. Note that the solution for $\sigma$ in the LDA is systematically below the one obtained from the boundary value problem, which agrees with the fact that radial gradients in AdS tend to favor $\chi$SB. Accordingly, the transition temperature is also systematically below in the LDA compared to the one obtained through the boundary value problem. 

\begin{figure}
    \centering
    \includegraphics[width=0.95\linewidth]{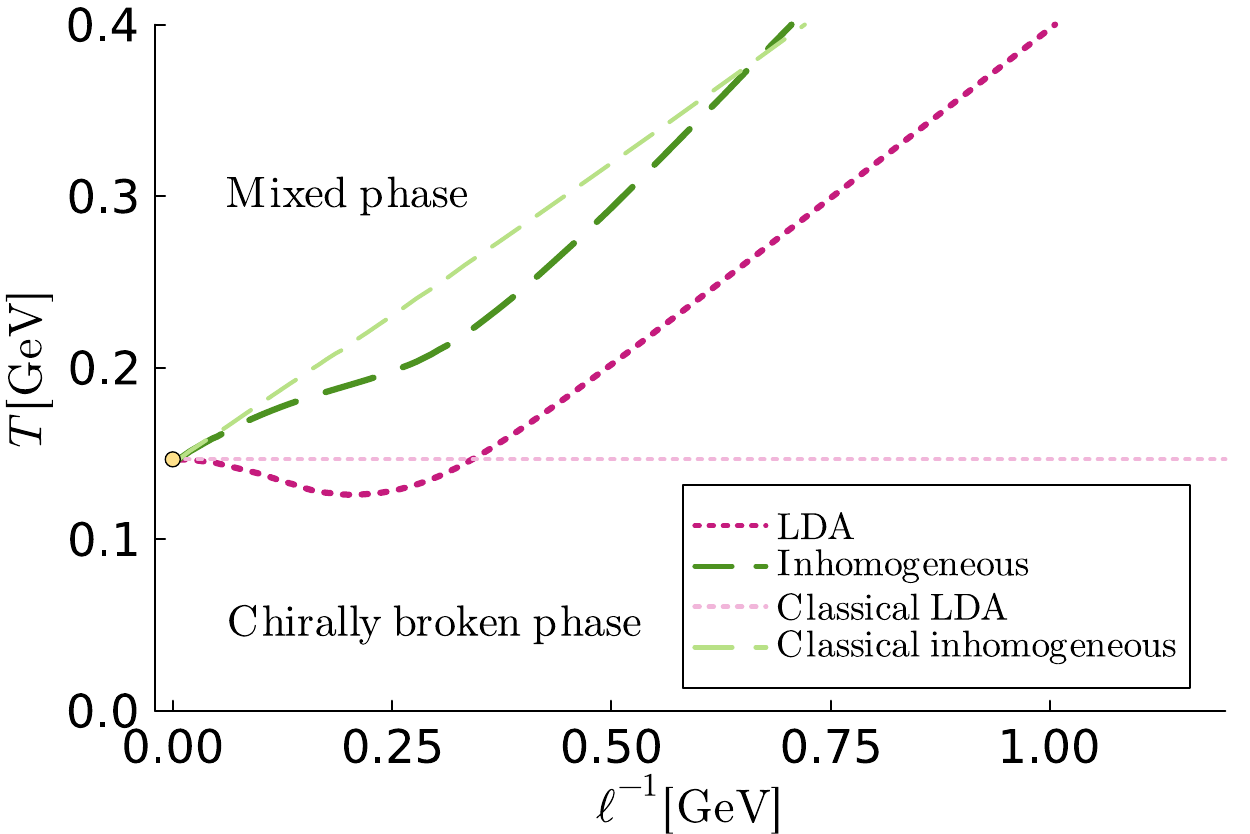}
    \caption{$T$-$\ell$ phase diagrams for $\chi$SB in AdS spacetime obtained in the mean-field approximation of the quark meson model. Dashed lines correspond to crossover transition temperatures obtained from the gap equation using the full quantum corrected condensate given in Eq. \eqref{eq_fermion_total}, while dotted lines are obtained using the classical version of the condensate given in Eq. \eqref{eq:fermion_cond_class}. Pink lines are obtained from the LDA while green lines are obtained from the solution to the boundary value problem described in Sec. \ref{sec:bdy_problem}.}
    \label{fig:phase_diag}
\end{figure}

In Fig. \ref{fig:phase_diag} (dashed lines) we show the phase diagram of this model. In the limit $\ell\to \infty$, the Ricci scalar vanishes and we recover the result from Minkowski spacetime that the critical temperature is about $T\simeq 150$ MeV both in the LDA and in the local inhomogeneous treatments. In the LDA, the critical temperature shows a non-monotonic trend: as the curvature of spacetime increases, the critical temperature decreases until it reaches a minimum value $T_c^{\rm min }\simeq 126$ MeV when $\ell\simeq 4.8$ GeV$^{-1}$. Further increasing the curvature we enter a regime in which the critical temperature scales linearly with the inverse of the AdS scale, $\ell^{-1}$. The critical temperature obtained from the solution of the boundary value problem is monotonically increasing, and systematically above the one obtained from the LDA. In the lower region of the diagram, the system is in a chirally broken phase, although the order parameter $\sigma$ can still be inhomogeneous, while in the upper part the system is considered to be in a mixed phase, with chiral symmetry restored as we approach the center of AdS. 

\section{Influence of the Hawking-Page transition}\label{sec:HP}

In Sec. \ref{sec:finite_effects} we have described $\chi$SB in a thermal AdS background. However, it is well-known that such thermal AdS background can be metastable, depending on the temperature $T$, decaying towards a Schwarzschild AdS black hole. This is known as the Hawking-Page transition \cite{Hawking:1982dh}, and is of first order nature. Specifically, such transition happens at a temperature 
\begin{equation}\label{eq:HP}
T_{\rm HP} = \frac{1}{\pi\ell}\,.    
\end{equation}
If the temperature is below $T_{\rm HP}$ then the thermal AdS background is thermodynamically favored, while if the temperature is above $T_{\rm HP}$ the AdS black hole background is preferred.

The results of Sec. \ref{sec:finite_effects} describe the phase of the system on top of the thermal AdS background, and the results hold in classical gravity. Alternatively, when the pure AdS phase is metastable, it should be understood that matter has thermalized in the thermal AdS background, and the decay towards the black hole is yet to happen. 

We would like to understand what are the expected changes of the phase diagram presented in Fig. \ref{fig:phase_diag} once the Hawking-Page transition is taken into account. Formally, this would require the evaluation of the fermion condensate of the AdS Schwarzschild background, and the techniques of Refs. \cite{Flachi:2011sx,Quinta:2019hrf,DeMott:2022qux} could be employed. The evaluation of the fermion condensate through the heat-kernel expansion employed in the aforementioned references is left for future work. 

Alternatively, we can obtain some insight on the phase of the system in the black hole background by using the expectation value of the fermion condensate computed \'a la Zubarev \cite{Zubarev:1979afm,Becattini:2012tc,vanWeert:1982}, that is to evaluate it assuming local thermal equilibrium (LTE), by which the temperature becomes an effective function of the spacetime coordinates. We shall refer to expectation values in the local thermal equilibrium as the classical expectation values. As we will see, the thermal contribution to the fermion condensate takes the same form as in flat space, albeit with an effective temperature that follows the Tolman-Ehrenfest law \cite{Tolman:1930ona,Tolman:1930zza}.

\subsection{Zubarev approach}

The thermal expectation values in local thermal equilibrium \cite{Becattini:2014yxa} are obtained, to leading order in gradients, with the following density operator:
\begin{equation}\label{eq:density}
    \hat\rho_{\rm LTE}(x) = e^{-\beta_\mu(x) \hat P^\mu}\,,
\end{equation}
where $\hat P_\mu$ is the momentum operator and $\beta_\mu(x) = u_\mu(x)/T(x)$ is the thermal four-velocity. The condition that there is local termal equilibrium is equivalent to $\beta_\mu$ being a time-like killing vector : $\nabla_{(\mu}\beta_{\nu)}=0$. We further demand that the four-velocity is normalized as $u_\mu u^\mu = -1\,$. The operator \eqref{eq:density} takes the same form as in flat space albeit with an effective temperature, and therefore the same logic applies for the expectation value of the fermion condensate. In this formalism, the classical expectation value of the fermion condensate for $N_f$ number of flavors, $N_c$ number of colors, and fermion mass equal to $g\sigma(x)$, is given by the same formula as in flat space with the replacement $T\to T(x)$:
\begin{equation}\label{eq:fermion_cond_class}
    \langle \overline \psi \psi\rangle_{\rm class.}(x) = g\sigma(x)\dfrac{2 N_fN_c}{\pi^2}\int_{g\sigma(x)}^\infty dE\dfrac{\sqrt{E^2-[g\sigma(x)]^2}}{e^{E/T(x)} + 1}\,,    
\end{equation}
 where the effective temperature $T(x)$ is obtained as a solution of the Killing equation for $\beta_\mu$ for a given four velocity $u^\mu$. Note that the vacuum contribution of the fermion condensate has been renormalized onto the parameters of the model. This is in contrast to the quantum fermion condensate \eqref{eq_fermion_total} on the AdS background, where the vacuum contribution is necessary to reproduce the flat space ($\ell\to\infty$) result for the fermion condensate.
 
\subsubsection{Thermal AdS background}\label{sec:Zubarev_AdS}
 
For the AdS metric given in Eq. \eqref{eq:AdS_metric}, the normalised four-velocity is chosen as
 \begin{equation}
     u_{\rm AdS} =u_{\rm AdS}^\mu\partial_\mu= \dfrac{1}{\ell}\cos r \partial_t\,.
 \end{equation}
Assuming that the effective temperature depends only on the AdS radial coordinate $r$, the set of Killing equations for $\beta^\mu$ has a single non-trivial component:
\begin{equation}
    \nabla_{(t} \beta_{r)} = \dfrac{\ell }{T(r)^2\cos r}\left(T(r)\tan r + \partial_r T(r)\right) =0\,.
\end{equation}
As a result, the effective temperature in an AdS spacetime is given by
\begin{equation}\label{eq:TE_AdS}
    T^{\rm AdS}(r) = T_0 \cos r\,,
\end{equation}
where $T_0$ is an integration constant. Note that the temperature vanishes at the boundary $r=\pi/2$, which is in line with the vanishing of the thermal contribution to the quantum expectation value of the fermionic condensate in Eq. \eqref{eq:scalar_cond}. Therefore, the boundary value of the condensate will always be $\sigma=f_\pi$, and the Zubarev approach reproduces the fact that chiral symmetry is broken close to the boundary of AdS.

\subsubsection{Schwarzschild-AdS background}
 
Similarly, we can obtain the effective temperature in the Schwarzschild-AdS black hole background. In this case, we can write the metric as 
\begin{equation}\label{eq:metric_BH}
    ds^2 = -f(\rho)dt^2 + \dfrac{d\rho^2}{f(\rho)} + \rho^2 d\Omega_2^2\,,
\end{equation}
where $d\Omega_2^2$ is the line element of the unit 2-sphere, $\rho$ is the radial coordinate, and the function $f(\rho)$ is explicitly given by
\begin{equation}
    f(\rho) = \dfrac{\rho^2}{\ell^2}+1-\dfrac{m}{\rho}\,.
\end{equation}
The boundary of spacetime is located at $\rho\to \infty$, where the metric becomes asymptotically AdS. The black hole event horizon is located at the outermost zero of $f$, that is 
\begin{equation}
    \rho_h^3  + \ell^2 \rho_h - m \ell^2  =0\,.
\end{equation}
As usual, the Hawking temperature $T_H=\kappa/(2\pi)$ of the black hole can be determined from the surface gravity $\kappa$,
\begin{multline}
    \kappa^2 = \lim_{\rho\to \rho_h}\left(-\dfrac{1}{2}\nabla_\mu k_\nu \nabla
    ^\mu k^\nu\right) = \lim_{\rho\to \rho_h}\dfrac{1}{4}(\partial_\rho f)^2 \\= \dfrac{1}{4 \rho_h^2}\left(1+3 \dfrac{\rho_h^2}{\ell^2}\right)^2\,,
\end{multline}
where $k = k^\mu\partial_\mu = \partial_t$ is a timelike killing vector and $\nabla_\mu k_\nu$ has the non-trivial components
\begin{equation}
    \nabla_tk_\rho = -\dfrac{1}{2}\partial_\rho f\,,\qquad \nabla_\rho k_t = \dfrac{1}{2}\partial_\rho f\,.
\end{equation}
Consequently, the Hawking temperature for the AdS Schwarzschild black hole is given by
\begin{equation}
    T_H = \dfrac{1}{4\pi \rho_h}\left(1+3\dfrac{\rho_h^2}{\ell^2}\right)\,.
\end{equation}

In order to obtain the effective temperature stemming from the local thermal equlibrium configuration, we consider the normalised four-velocity
\begin{equation}
    u_{\rm BH} = u_{\rm BH}^\mu\partial_\mu=\dfrac{1}{\sqrt{f(\rho)}}\partial_t\,, \qquad u^\mu_{\rm BH} u_{\mu}^{\rm BH} = -1\,.
\end{equation}
Then, we define the thermal four-velocity $\beta^\mu_{\rm BH} = u^\mu_{\rm BH}/T^{\rm BH}(x)$ and impose the killing equation for $\beta^\mu_{\rm BH}$ to find $T^{\rm BH}(x)$. Assuming that the effective temperature depends only on the radial coordinate $\rho$, the killing equation has a single non-trivial component,
\begin{equation}
    \nabla_{(t }\beta_{\rho)}=\dfrac{T(\rho)\partial_\rho f(\rho) + 2 f(\rho)\partial_\rho T(\rho)}{2T(\rho)^2\sqrt{f(\rho)}} =0\,.
\end{equation}
Its solution gives the effective temperature in the black hole background:
\begin{equation}\label{eq:TE_BH}
    T^{\rm BH}(\rho) = \dfrac{T_0}{\sqrt{f(\rho)}} = \dfrac{T_0}{\sqrt{\rho^2/\ell^2+1-m/\rho}}\,,
\end{equation}
where the integration constant is again denoted as $T_0$ and is identified with the Hawking temperature $T_H$\footnote{Recall that the Hawking temperature can be obtained demanding that the Euclidean version of the metric \eqref{eq:metric_BH} is regular at the horizon. On the other hand, a static observer measures temperature locally with its proper time $ds = \sqrt{f}d\tau$, where $\tau$ is the periodic Euclidean time with period $1/T_H$. As a consequence, the proper (imaginary) time is also periodic with period $\sqrt{f}/T_H\equiv1/T^{\rm BH}$ and we recover the Tolman-Ehrenfest law with the explicit identification of $T_0=T_H$.}. 

The effective temperature vanishes at the boundary ($\rho\to\infty$) while it diverges at the event horizon, where $f(\rho_h)=0$. Since the boundary is asymptotically AdS, and the effective temperature vanishes again at the boundary, then the value of the condensate $\sigma$ at the boundary will again be $\sigma(\rho\to\infty)=f_\pi$, and chiral symmetry is broken close to the boundary. 

The diverging temperature at the horizon generically results into a divergent classical fermion condensate \eqref{eq:fermion_cond_class}, unless $\sigma$ vanishes sufficiently fast also at the horizon. This feature is an artifact of the LTE approach, and a proper evaluation of the fermion condensate expectation value on the Hartle-Hawking state \cite{Hartle:1976tp} should remain finite also at the horizon (see Ref. \cite{Casals:2012es} for an example of regular expectation values on a Kerr black hole geometry).

\subsection{Classical phase diagram}

We shall employ the classical fermion condensate \eqref{eq:fermion_cond_class} as a source for the gap equation, which we shall refer to as the \textit{classical} gap equation, and discuss the resulting \textit{classical} phase diagram. We first compare the classical phase diagram in the AdS background with the phase diagram obtained in Sec. \ref{sec:finite_effects}. Later, we analyse the results for the AdS black hole an present a phase diagram where the first order HP transition is taken into account.

\subsubsection{Thermal AdS background}

We start outlining the results in the LDA, i.e. neglecting the gradients of $\sigma$ in the classical gap equation. As pointed out in Sec. \ref{sec:Zubarev_AdS}, chiral symmetry is always restored close to the boundary of AdS. We have numerically verified that the condensate $\sigma$ is a monotonous function of the AdS radial coordinate $r$. It then follows, as in Sec. \ref{sec:finite_effects}, that the inhomogeneous phase of the system can be characterized in terms of the value of the condensate $\sigma$ at the center of AdS.

According to the Tolman-Ehrenfest law in AdS \eqref{eq:TE_AdS}, the effective temperature at the center is just $T(r=0)=T_0$, and the classical gap equation at $r=0$ in the LDA is independent of the AdS scale $\ell$. Therefore, the gap equation at the center of AdS is just the Minkowski gap equation, and a crossover transition takes place at $T\simeq 150$ MeV for any value of the curvature. This transition line is depicted in Fig. \ref{fig:phase_diag} with label ``classical LDA". The major difference with the result for  Minkowski spacetime is that above the transition temperature, the system is in a inhomogeneous phase, chirally broken close to the boundary and restored close to the center.

We now reinstate the gradients of $\sigma$ and discuss the phase diagram that stems from the classical differential gap equation. The analysis of the differential equation for AdS given in Secs. \ref{sec:LI} was provided for a general fermion condensate $\langle\overline\psi\psi\rangle$, and the conclusions from such analysis can directly be used for the classical fermion condensate. In particular, the classical differential gap equation is to be solved demanding that $\sigma$ is regular at the center and at the boundary of AdS. Furthermore, the value of the condensate at the boundary is still the same as given in the LDA equation, which for the classical case is just $\sigma_{\rm bdy} = f_\pi$ for any temperature. 

We solve the differential classical gap equation and characterize the phase of the system by the value of $\sigma$ at the center of AdS. The resulting phase diagram is also presented in Fig. \ref{fig:phase_diag} with the transition line labeled as ``classical inhomogeneous".

Analyzing all the information given in Fig. \ref{fig:phase_diag}, we conclude that the Zubarev approach is able to reproduce the order of the phase transition and the value of the transition temperature only for weakly curved spacetimes $\ell\gg1$. In particular, it seems to reproduce the slope of the transition curve at $\ell\to\infty$. Generic features of the inhomogeneous phases are also reproduced satisfactorily, as the fact that chiral symmetry always remains spontaneously broken close to the boundary of AdS.

\subsubsection{Schwarzschild-AdS background}

In this case, the result for the LDA can be obtained automatically by inspection of the classical fermion condensate with the effective temperature for the black hole AdS background. The effective temperature \eqref{eq:TE_BH} vanishes at the boundary and diverges at the black hole event horizon. As a consequence, the boundary value of the condensate is just $\sigma_{\rm bdy}=f_\pi$ and chiral symmetry is again broken close to the boundary of AdS. At the horizon, $\rho=\rho_h$, the divergent temperature in the algebraic gap equation can only be compensated by a vanishing condensate: $\sigma_{\rho=\rho_h}=0$. It then follows that the system is always in a inhomogeneous phase when considered in the Schwarzschild-AdS background.

We now turn our attention to the classical differential gap equation in the black hole background \eqref{eq:metric_BH}. In this case, the classical gap equation is given by 

\begin{equation}\label{eq:gap_diff_BH}
    f\partial_\rho^2\sigma + \left(\partial_\rho f + \frac{2}{\rho}f\right)\partial_\rho\sigma - \lambda(\sigma^2-v^2)\sigma=g\langle\overline\psi \psi\rangle_{\rm class.}-h
\end{equation}
where the condensate is given by \eqref{eq:fermion_cond_class} with the effective temperature \eqref{eq:TE_BH}. In order to determine the regular solutions to \eqref{eq:gap_diff_BH}, we first obtain the asymptotic solution around the its singular points, which in this case are the boundary, $\rho\to\infty$, and the horizon, $\rho\to\rho_h$. The spacetime \eqref{eq:metric_BH} is asymptotically AdS, and therefore the near boundary analysis of Sec. \ref{sec:LI} again applies, and we recover the \textit{motto} of this work: chiral symmetry is always broken close to the boundary. The asymptotic solution close to the horizon needs to be worked out separately and we present it below. 

As a preliminary step to solving Eq. \eqref{eq:gap_diff_BH} around the horizon $\rho_h$, we expand the classical condensate \eqref{eq:fermion_cond_class} on the black hole background around $\rho\to\rho_h$. Since the effective temperature \eqref{eq:TE_BH} diverges at the horizon, we change variables in the integral, denoting $\mathbf{x} = E/T^{\rm BH} = E \sqrt{f}/T_H$, and expand around $T\to \infty$. Then, the classical condensate is expanded as
\begin{align}\label{eq:ferm_exp_bh}
    &\langle\overline\psi\psi\rangle_{\rm cl.}(\rho) =\nonumber\\ &=T_{\rm BH}^3 \left(\frac{g \sigma}{T_{\rm BH}}\right)\frac{2 N_f N_c}{\pi^2}\int_0^\infty d\mathbf{x}\dfrac{\mathbf{x}}{e^{\mathbf{x}}+1} + O\left(\left[\frac{g\sigma}{T_{\rm BH}}\right]^3\right)\nonumber\\ & = \frac{ N_f N_c }{6} g\sigma(\rho) \frac{T_H^2}{f(\rho)}  + O\left(\left[\frac{g\sigma}{T_{\rm BH}}\right]^3\right)\,,
\end{align}
where we have assumed that the effective fermion mass $g\sigma$ does not diverge at the event horizon. Note that at the horizon, the function $f(\rho)$ vanishes, and the fermion condensate may or may not diverge depending on the behaviour os $\sigma$ at $\rho_h$. We propose a Frobenius ansatz for the condensate $\sigma$ around the critical point:
\begin{equation}\label{eq:Frob_bh}
    \sigma = (\rho - \rho_h)^k\sum\sigma^h_n (\rho-\rho_h)^n + \dots
\end{equation}
where the ellipsis contain possible subleading terms to the Frobenius ansatz. 
The assumption that $\sigma$ does not diverge at the horizon translates into the condition $k\geq0$. We substitute ansatz for $\sigma$ \eqref{eq:Frob_bh} and the expansion for the fermion condensate \eqref{eq:ferm_exp_bh} into the differential gap equation \eqref{eq:gap_diff_BH}, and expand the latter as $\rho\to\rho_h$. It is straightforward to verify that the leading contributions to  \eqref{eq:gap_diff_BH} are
\begin{equation}\label{eq:pertlead}
    h + (\rho-\rho_h)^{k-1}\dfrac{\sigma_0^h T_H}{24\pi }(96k^2 \pi^2  - g^2 N_fN_c)+\dots=0\,,
\end{equation}
where the ellipsis contains terms that are subleading to $h$ or to $(\rho-\rho_h)^{k-1}$. Which of the two terms in Eq. \eqref{eq:pertlead} is actually the leading term depends on the value of $k$ and we have three possibilities:
\begin{itemize}
    \item $0<k<1$

    Then the second term in Eq. \eqref{eq:pertlead} is leading, and a non-trivial solution requires that the Frobenius exponent is 
    \begin{equation}\label{eq:exp}
        k = \frac{g}{4\pi}\sqrt{\frac{N_f N_c}{6}}\simeq 0.26
    \end{equation}
    where we have taken the positive solution and substituted the parameters of the model. Indeed, the obtained value for $k$ lies in the range $[0,1]$, and the solution is consistent with the assumption for this case. We can perturbatively obtain the first few coefficients of the expansion \eqref{eq:Frob_bh}. We find that $\sigma_0^h$ is an undetermined integration constant, while $\sigma_1^h$ is given by
    \begin{equation}\label{eq:s1h}
        \sigma_1^h = \dfrac{g^2 + 2 g \pi(1-8\pi\rho_h T_H)-8\pi^2\rho_h^2v^2\lambda}{16\pi^2(g+2\pi)\rho_h^2 T_H}\sigma_0^h\,.
    \end{equation}
    Note that for non-integer $k$, the ansatz \eqref{eq:Frob_bh} is not able to compensate for the term proportional to $h$. In this case, the ellipsis in Eq. \eqref{eq:Frob_bh} contains a regular expansion. In particular, we find
    \begin{align}\label{eq:sigma_hor}
        &\sigma = (\rho-\rho_h)^{k}\left[\sigma_0^h + \sigma_1^h(\rho-\rho_h) \right]  -\dfrac{h(\rho-\rho_h)}{4\pi T_H(1-k^2)} \nonumber\\&-h \dfrac{2+k^2-\rho_h(8\pi T_H+\rho_h v^2\lambda)}{16\pi^2T_H^2(1-k^2)(4-k^2)}(\rho-\rho_h)^2\nonumber\\& + O([\rho-\rho_h]^3)+O([\rho-\rho_h]^{k+2}) +O([\rho-\rho_h]^{nk}) 
    \end{align}
    where the coefficient $\sigma_1^h$ is given in Eq. \eqref{eq:s1h} and the exponent $k$ in Eq. \eqref{eq:exp}. Note that subleading terms can appear with generic powers of $k$, i.e. of the form $(\rho-\rho_h)^{nk}$ for integer $n$. This is a consequence of the non-linear nature of the differential equation Eq. \eqref{eq:gap_diff_BH}. 

    The asymptotic solution around the event horizon has a single integration constant\footnote{The second integration constant would be proportional to the negative solution to Eq. \eqref{eq:pertlead}, i.e. $\sigma\sim C (\rho-\rho_h)^{-|k|}$. Such contribution is divergent and therefore it is necessary that $C=0$.}, denoted $\sigma_0^h$. It follows from Eq. \eqref{eq:sigma_hor} that the condensate $\sigma$ vanishes at the black hole event horizon: $\sigma(\rho_h)=0$.   

    Finally, we would like to point out that the fermion condensate at the event horizon \eqref{eq:ferm_exp_bh} diverges as $(\rho-\rho_h)^{k-1}$ in spite of the condensate $\sigma$ being finite everywhere. This divergence is compensated in Eq. \eqref{eq:gap_diff_BH} through the gradient terms. 

    \item $k=1$

    In this case both terms in Eq. \eqref{eq:pertlead} are of the same order. Then, the solution to the leading term is given by 
    \begin{equation}
        \sigma_0^h = -\dfrac{h}{4\pi T_H\left(1-\frac{g^2}{16\pi^2}\frac{N_fN_c}{6}\right)}\,.
    \end{equation}
    The next coefficients in the expansion for $\sigma$ \eqref{eq:Frob_bh} can again be obtained in a perturbative fashion, solving Eq. \eqref{eq:gap_diff_BH} around $\rho\to\rho_h$. The next coefficient is given by
    \begin{equation}
        \sigma_1^h = -h \dfrac{2+\frac{g^2}{16\pi^2}\frac{N_fN_c}{6}-\rho_h(8\pi T_H+\rho_h v^2\lambda)}{16\pi^2T_H^2\left(1-\frac{g^2}{16\pi^2}\frac{N_fN_c}{6}\right)\left(4-\frac{g^2}{16\pi^2}\frac{N_fN_c}{6}\right)}\,.
    \end{equation}
    The branch of solutions for $k=1$ has no integration constant, and is similar in structure to the previous solution (with $0<k<1$) once the integration constant of the previous branch is set to $0$. 

    Note that the condensate $\sigma$ also vanishes at the horizon for this branch of solutions. In this case, the fermion condensate \eqref{eq:ferm_exp_bh} does not diverge, and reaches a constant value at the event horizon. 
    \item $k>1$
    
    In this case, the term proportional to $h$ is leading in Eq. \eqref{eq:gap_diff_BH} and there exist no solution for such values of $k$.
    
\end{itemize}

While the analysis of the differential equation \eqref{eq:gap_diff_BH} is slightly more involved than its analogue in the LDA, the conclusion remains that the condensate $\sigma$ vanishes at the black hole event horizon. We are then led to conclude that chiral symmetry gets restored in a neighborhood of the black hole. The conclusion that the condensate $\sigma$ vanishes at the horizon is in contrast with the results reported in Ref. \cite{Tanaka:2026geo}, where non-zero values of $\sigma$ seem to be possible also near the horizon. The underlying reason for this difference is that the classical condensate \eqref{eq:fermion_cond_class} diverges at the horizon, and a regular solution to the differential or algebraic gap equation must result into $\sigma(\rho_h)=0$. We expect this feature to change, and a richer phase structure around the black hole to appear if the fermion condensate or thermodynamic potential are evaluated in the Hartle-Hawking state or with the methods of Ref. \cite{Flachi:2011sx}.

All in all, we have shown that the classical system is always in a chirally broken phase close to the boundary, while it is always in the restored phase close to the horizon. Therefore, the system is in a mixed phase in the black hole AdS background regardless of the values of temperature $T_H$ and AdS scale $\ell$.

\subsubsection{Final discussion}

\begin{figure}
    \centering
    \includegraphics[width=0.95\linewidth]{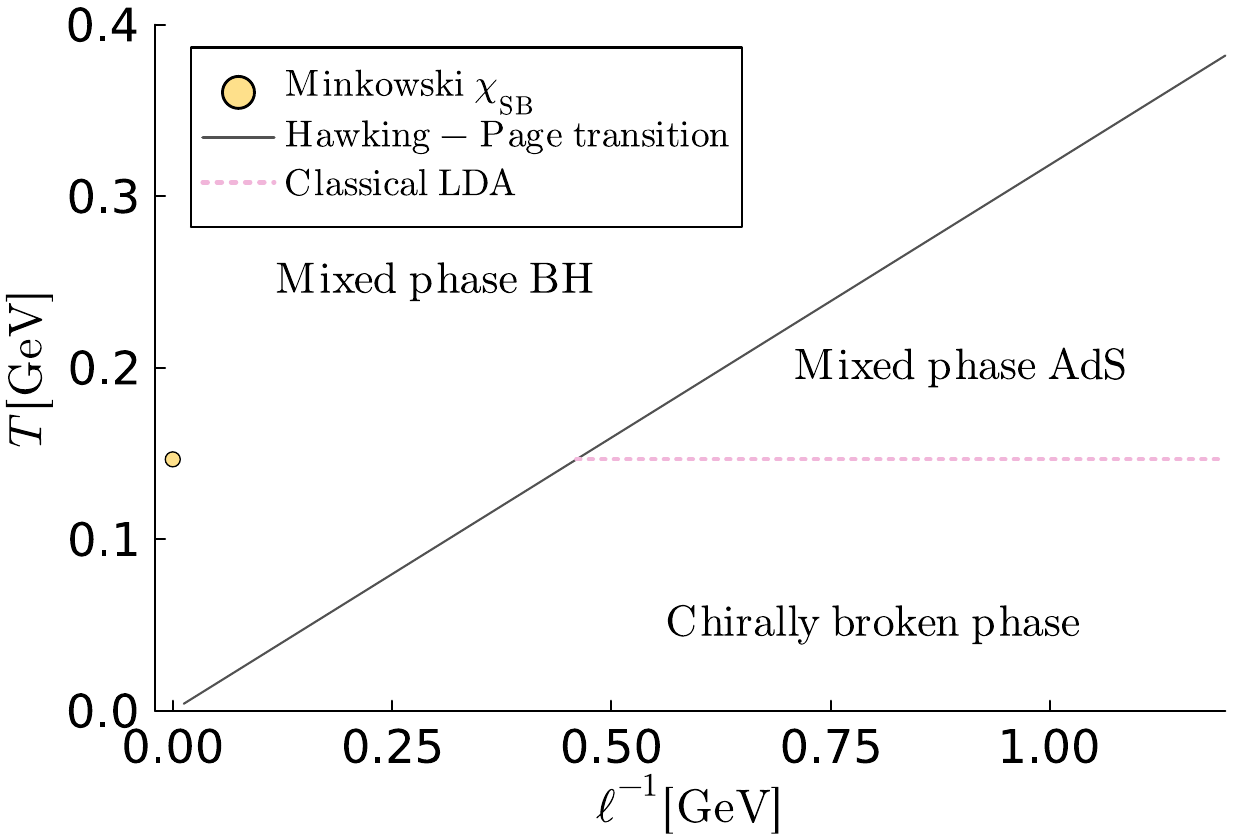}
    \caption{Phase diagram of the quark meson model in a background spacetime with an asymptotic AdS boundary in the local thermal equilibirum approximation. In the upper half of the phase diagram, the black hole background is dominant (except at zero curvature), and chiral symmetry is restored at the boundary and broken at the horizon. In the lower half, the AdS background is thermodynamically favored and the classical phase structure of Fig. \ref{fig:phase_diag} applies.}
    \label{fig:phase_diag_BH}
\end{figure}

We conclude this section combining the classical results for the phase diagram obtained for the AdS and Schwarzschild-AdS backgrounds together with the first order Hawking-Page transition phase transition taking place at $T_{\rm HP}$ of Eq. \eqref{eq:HP}. The resulting phase diagram is presented in Fig. \ref{fig:phase_diag_BH}. In the upper half of the diagram, that is for $\pi T>\ell^{-1}$, the black hole background is preferred and the classical description predicts that the system is always in a mixed phase, with chiral symmetry restored at the horizon and broken at the boundary. Note the exception of vanishing curvature $\ell^{-1}\to0$, corresponding to Minkowski space, where the black hole configuration is not stable and the well-known crossover transition takes place at $T\simeq 150$ MeV. In the lower half of the diagram, the AdS background is thermodynamically favoured and the phase of the system has been described already in Fig. \ref{fig:phase_diag}. The LDA predicts a phase transition from a purely chirally broken phase to a mixed phase at $T\simeq 150$ MeV. Including the gradients of $\sigma$, such transition is only possible at higher temperatures, where the AdS background is no longer stable. As a consequence, the solutions to the differential gap equation predict that chiral symmetry is always in the pure chirally broken phase when the AdS background is preferred.

\section{On the holographic correspondence}\label{sec:holo}

The formulation of the quark-meson model in AdS spacetime may be viewed through the lens of the holographic correspondence. In this case, the theory considered in this work would be dual to a $2+1$-dimensional quantum field theory with a scalar operator, which we denote $\Sigma$, dual to the condensate $\sigma$. The path integral over the fermion fields would modify the potential for $\sigma$, introducing some explicit dependence on the holographic radial coordinate. Such explicit coordinate dependence typically appears as a result of writing the action for the model partially on-shell, as exemplified in Ref. \cite{Gauntlett:2018vhk}.

The asymptotic solution of the Klein-Gordon equation for $\sigma$ close to the boundary of AdS, given in Eqs. \eqref{eq:Taylor_bdy_fermion} and \eqref{eq:dsigma}, connects with the standard holographic dictionary. The integration constants $\mathcal{C}_-$ and $\mathcal{C}_+$ correspond to the source and the expectation value of $\Sigma$ respectively, while the exponent $\Delta_+$ in \eqref{eq:dsigma} is the conformal dimension of the operator $\Sigma$. It should be noted from Fig. \ref{fig:m_eff_Deltas} that the exponent $\Delta_-$ is negative. Therefore, a nonzero source $\mathcal{C}_-$ would yield an infinite contribution of the condensate $\sigma$ close to the boundary, which would be inconsistent with the asymptotic AdS structure. It is then required that $\mathcal{C}_-=0$. The previous observation is closely tied to the fact that $\Delta_+>d=3$, so the operator dual to $\sigma$ is non-renormalizable in the dual field theory and should not be sourced. 

The scalar field $\sigma$ can acquire a non-trivial profile in the holographic radial coordinate despite the fact that the source is set to zero. This scenario would correspond to a holographic renormalization group flow that is driven the vacuum expectation value of the scalar operator $\langle\Sigma\rangle$. At zero temperature, the condensate $\sigma$ is constant and therefore $\mathcal{C}_+=0$, while as temperature is increased, we must have $\mathcal{C}_+\neq0$. Therefore, the mixed phase and chirally broken phase in the AdS background can be related to two thermodynamic phases in the dual QFT that are distinguished by $\langle \Sigma\rangle$. Finally, the Hawking-Page phase transition has been argued to correspond holographically to a confinement/deconfinement phase transition \cite{Witten:1998zw}.

\section{Conclusions \& Discussion} \label{sec:conc}

We have studied $\chi$SB using the quark meson model in a thermal AdS background. We have generically found that quark matter in AdS is either in a chirally broken phase throughout, or in a mixed phase where chiral symmetry is broken close to the boundary and restored at the center. The effective fermion mass is sensitive to the curvature effects via the fermion condensate \eqref{eq_fermion_total} and through the gradient terms in the Klein-Gordon equation \eqref{eq:gap_difeq_AdS}. As shown in Fig. \ref{fig:phase_diag}, the effects of the former are non-universal, slightly favoring the restored phase for small curvatures and preventing it at large curvatures. The gradient terms, on the other hand, always favor the spontaneous breaking of chiral symmetry. These results are consistent with the previously observed trend that negatively curved spaces inhibit the restoration of chiral symmetry. 

Unlike (negative) curvature, temperature tends to restore chiral symmetry, as it happens in flat space. Interrestingly, we have found that there is no temperature at which the system is in the chirally restored phase everywhere. In fact, chiral symmetry is always broken close to the boundary of AdS, where the thermal part of the condensate \eqref{eq:scalar_cond} vanishes. 

The analysis of this work has been carried out in the probe limit, i.e.  neglecting the backreaction of the matter fields onto the geometry. Initially, we studied the system on an AdS background, where the expectation value of the fermion condensate was expressed in closed form in \cite{Ambrus:2017cow}. However, even in the probe limit, a first order phase transition takes place that modifies the background geometry: the Hawking-Page phase transition. At low temperatures, the thermal AdS background is thermodynamically stable. As the temperature increases, the background becomes metastable, and the black hole AdS geometry becomes the dominant saddle in the gravitational path integral. This transition happens at a critical temperature $T_{\rm HP}$ \eqref{eq:HP}. We then proceed to study $\chi$SB on a Schwarzschild-AdS background. We evaluate the fermion condensate in the black hole background is done via the Tolman-Ehrenfest law. We verify that this approach yields qualitative agreement with the results of Sec. \ref{sec:finite_effects} in the AdS background, where the fermion condensate of Ref. \cite{Ambrus:2017cow} is employed. Through the Tolman-Ehrenfest law, we have found that chiral symmetry is always restored close to the event horizon. We attribute this phenomenon to the diverging effective temperature at the horizon, which causes the fermion condensate to melt. We expect a rigorous evaluation of the fermion condensate in a black hole background to reveal a richer phase structure close to the event horizon, as found in Ref. \cite{Tanaka:2026geo} for black holes in flat space. Finally, we find that chiral symmetry is generically broken close to the boundary, even in the black hole background.

In the context of this work, it would be interesting to address several open questions:
\begin{itemize}
    \item The evaluation of the fermion condensate in the AdS background, as described in Ref. \cite{Ambrus:2017cow} assumes a constant fermion mass. This is in contrast to the fact that the effective fermion mass is dynamically generated in the theory, and usually becomes inhomogeneous in a curved background. In this work, the fermion condensate of Ref. \cite{Ambrus:2017cow} has been used as a proxy in the gap equation. While we expect the result at low curvatures to be reasonably accurate, the evaluation of the condensate with an inhomogeneous mass could modify the phase diagram qualitatively at large curvatures. 
    \item Similarly, the phase structure of the theory in the black hole background is heavily affected by the effective divergent temperature in the local thermal equilibrium approach. A rigorous treatment of the problem requires the use of the fermion condensate in an AdS-Schwarzschild spacetime. It is likely that a much richer phase structure would arise in this case.
    \item The study of chiral symmetry in AdS space is special, compared to the studies in Minkowski or de Sitter spaces, because of the first order Hawking-Page transition. This phase transition is also affected by the matter fields present in the theory. It would be interesting to characterize how the backreaction of the matter fields onto the geometry modifies the structure of the theory. This would require the knowledge of the fermion condensate and energy momentum tensor in a generic curved background and inhomogeneous mass. A first step in this direction has been taken in Ref. \cite{Thompson:2024akz}, where the semiclassical modification to AdS space due to quantized free fermions was addressed.
\end{itemize}

The previous three problems require the evaluation of the effective mesonic potential in a generic curved background with inhomogeneous mass, and the techniques of Ref. \cite{Flachi:2011sx} could be employed.

\begin{itemize}
    \item The fermion condensate in AdS is known not only at finite temperature, but also at finite angular velocity \cite{Ambrus:2021eod}. Naive rotation typically conficts with relativistic \textit{unbounded} systems in Minkowski spacetime due to the superluminal velocities that are reached beyond the light cylinder. A well-defined approach is to solve the Dirac equation in a bounded system with appropriate boundary conditions to prevent modes from extending beyond the domain \cite{Ambrus:2015lfr}. It is surprising that the results for chiral symmetry with rotation in the unbounded system are physically reasonable and comparable to those of the formally bounded construction, see Ref. \cite{Morales-Tejera:2025qvh} and references therein. 
    In this context, anti-de Sitter space is interesting because the unbounded system is also well defined provided that $\Omega\ell<1$. Therefore, studying $\chi$SB with rotation in AdS is a suitable way to assess the sanity of the predictions for chiral symmetry in unbounded flat space with rotation.
\end{itemize}

\acknowledgments 

I would like to thank P. Singha, C. Rosen, M. Chernodub and V. Ambrus for useful discussions in the development of this manuscript. This work was funded by the EU’s NextGenerationEU instrument through the National Recovery and Resilience Plan of Romania - Pillar III-C9-I8, managed by the Ministry of Research, Innovation and Digitization, within the project entitled ``Facets of Rotating Quark-Gluon Plasma'' (FORQ), contract no. 760079/23.05.2023 code CF 103/15.11.2022.

\bibliography{plasma}

\end{document}